\RequirePackage{fix-cm}
\documentclass[smallextended]{svjour3}       
\smartqed  
\usepackage{graphicx}
\usepackage{mathtools}
\usepackage{axodraw2}
\usepackage{tikz}
\usepackage{tikz-feynman}
\usepackage{amscd}
\usepackage{amsmath}
\usepackage{slashed}
\usepackage{gensymb}
\usepackage{amssymb}
\usepackage{bm}
\usepackage[normalem]{ulem} 
\journalname{Eur.\ Phys.\ J.\ C}
\begin{document}

\title{Constraints on the malaphoric $B_3-L_2$ model from di-lepton
  resonance searches at the LHC\thanks{This work was partially supported by STFC HEP Consolidated grants
ST/T000694/1 and ST/X000664/1.}
}

\titlerunning{Constraints on the malaphoric $B_3-L_2$ model}        

\author{Ben Allanach}


\institute{Ben Allanach \at
              DAMTP, University of Cambridge, Wilberforce Road, Cambridge, 
  CB3 0WA, United Kingdom \\
              \email{ben.allanach.work@gmail.com}
}

\date{Received: date / Accepted: date}

\maketitle

\begin{abstract}
We confront the malaphoric $B_3-L_2$ model with bounds coming from
  a search for resonances in the di-lepton channels at the 13~TeV LHC\@.
  In contrast to the original   $B_3-L_2$ model, 
  the $Z^\prime$ of the \emph{malaphoric} $B_3-L_2$ model has sizeable
  couplings to the lighter two 
  families; these originate from order unity kinetic mixing
  with the hypercharge gauge
  boson and ameliorate the fit to
  lepton flavour universality measurements in $B-$meson decays.
  The $Z^\prime$ coupling to the first two families of quark means that
  the resulting constraints from resonant di-lepton 
  searches are stronger.
  Nevertheless, we find that  for $M_{Z^\prime}>2.8$ TeV there remains
  a non-negligible region of allowed parameter
  space where the model significantly improves upon several
  Standard Model predictions for observables involving the $b \rightarrow s
  l^+ l^-$   transition. We estimate that the 3000
  fb$^{-1}$ HL-LHC will extend this sensitivity to 
  $M_{Z^\prime}= 4.2$ TeV.
\end{abstract}

\section{Introduction \label{sec:intro}}

The malaphoric $B_3-L_2$ model was introduced in Ref.~\cite{Allanach:2024ozu} in order to
ameliorate fits of the Standard Model (SM) to current $B$ meson decay observables
involving the $\bar b s \mu^+ \mu^-$ and $\bar b s e^+ e^-$ effective
couplings (and their CP
conjugates\footnote{Throughout the rest of this article, CP conjugates are to be
  understood as to be added implicitly to the effective vertices we mention.}). There are significant theoretical uncertainties in the
predictions of many of these observations, notably from the contribution from
non-local hadronic 
matrix elements.
The largest theory uncertainty, estimated from a combination of data and theory
in Ref.~\cite{Gubernari:2022hxn}, is from the charm loop contribution. There
are concerns that even the state-of-the-art calculations do not completely
capture all of the contributions; in particular loop diagrams involving
effective hadrons running in the loop~\cite{Ciuchini:2022wbq}.
Although no
estimate of all such 
contributions has been yet completed,
recent estimates~\cite{Gubernari:2022hxn,Mutke:2024tww,Isidori:2024lng} have 
bounded various different
contributions to it, finding that they are
too small to explain the current disparity between measurements and
state-of-the-art SM predictions. It is fair to say, however~\cite{Ciuchini:2022wbq}, that the
SM 
predictions for such observables require further work before evidence for new
physics can be unambiguously claimed. We shall investigate the case that the
additional effective hadronic contributions are small and fit a new physics
contribution to the measurements. 

One such new physics model that explains this new physics contribution is the
malaphoric $B_3-L_2$ model. 
The malaphoric $B_3-L_2$ model is a deformation of the original $B_3-L_2$ model~\cite{Bonilla:2017lsq,Alonso:2017uky,Allanach:2020kss}
to allow for sizeable kinetic mixing. In either $B_3-L_2$ model, an extra
spontaneously broken 
$U(1)_X$ gauge group is introduced in a direct product with the SM gauge
group. The $X$ charges of the SM fermions are assumed to be third family
baryon number minus second family lepton number, and the model requires the
addition of three families of SM-singlet right-handed Weyl fermionic fields
(right-handed neutrinos) for anomaly cancellation. Right-handed neutrinos also
provide a simple means of obtaining tiny neutrino masses via the seesaw
mechanism. In contrast to the SM gauge group, the $B_3-L_2$ gauge group
distinguishes between the different families. A SM-singlet complex scalar
field 
$\theta$ (the `flavon') that has a non-zero $X$ charge $X_\theta$ is assumed
to acquire a non-zero vacuum expectation value $\langle \theta \rangle \sim
{\mathcal O(1)}$ TeV and spontaneously break $U(1)_X$. 
Assuming that the SM Higgs doublet field has a zero $X$ charge leads to an
attractive property of the model: all off-diagonal CKM matrix
elements are predicted to be much smaller than unity aside from $|V_{us}|$ and
$|V_{cd}|$, agreeing with
measurements~\cite{ParticleDataGroup:2024cfk}. 
With the 
$U(1)_X$ gauge symmetry spontaneously broken, a massive spin-1 vector boson field $X^\mu$ remains, whose mass parameter
\begin{equation}
  M_X= \frac{g_X}{\sqrt{2}} X_\theta \langle \theta \rangle
\end{equation}  
is expected to be around the TeV scale. This electrically-neutral and
colourless  spin-1 bosonic field (the associated
particle is dubbed a 
$Z^\prime$ boson) has
tree-level flavour changing neutral 
currents. However, if one assumes that these are mostly associated with
the transitions of the up quarks and of neutrinos, they
are not subject to constraints on new physics coming from
tree-level flavour changing
currents between the $d$ and $s$ quark fields, which are extremely tight and
derive from measurements of 
$K-\bar K$ mixing.
In the years leading up to December 2022, new physics
effects
in the $\bar b s \mu^+ \mu^-$ effective coupling
provided a much better fit to $B-$meson decay data than did the SM\@.
In the $B_3-L_2$
model, this effective coupling is mediated at tree-level by the $Z^\prime$
boson as well as by loops involving electroweak gauge bosons.
Since December 2022 though, analyses from the LHCb collaboration indicate 
that a better fit results when there is a new physics contribution to the
$\bar b s e^+ e^-$ effective coupling in addition to that of $\bar b s \mu^+
\mu^-$~\cite{LHCb:2022qnv}.

The \emph{malaphoric} $B_3-L_2$ model achieves this by introducing a
substantial kinetic mixing between the hypercharge gauge boson field $B^\mu$ and the $X^\mu$
gauge boson; such a term is invariant under all of the symmetries in the model.
Since $B^\mu$ has interactions with both di-electron pairs and
di-muon pairs, the kinetic mixing term results in the $Z^\prime$ boson also
having 
these interactions. Ref.~\cite{Allanach:2024ozu} demonstrates that this results in a
significantly better global fit to $b\rightarrow s l^+ l^-$ 
transition measurements\footnote{In the present paper, $l^\pm \in
\{e^\pm,\ \mu^\pm \}$.}, than does the original model. Ref.~\cite{Allanach:2024ozu}
used the language of deriving SMEFT coefficients from integrating out the
$Z^\prime$. This was appropriate because all of the constraints applied were
on measurements at energy scales much less than the mass of the $Z^\prime$,
$M_{Z^\prime}$. Here, we shall need a different treatment because we will
(in addition) apply constraints coming from LHC resonance
searches. These necessarily involve observables measured at a scale of the $M_X$ and so the
SMEFT cannot be used to characterise them.

We shall therefore introduce the malaphoric $B_3-L_2$ model in the effective
field theory that \emph{includes} the associated $Z^\prime$, rather than
having it integrated out of the theory as in the SMEFT characterisation. Our
discussion 
roughly follows (and agrees with) Refs.~\cite{Babu:1997st,Cheng:2024hvq}, but has a more compact notation. 
We have three gauge fields which are electrically neutral and which mix. In the model
gauge eigenbasis, these are put into a 3-vector (vectors will be written here in
bold typeface) ${\bf \hat G}_\mu:=(\hat B_\mu,\ \hat W^3_\mu,\ \hat X_\mu)^T$, where the
circumflex denotes that the field is in the original gauge kinetically-mixed
basis and $\hat W^3_\mu$ 
is the electrically neutral $SU(2)$ gauge field. 
We begin with the relevant important terms of the Lagrangian density involving
${\bf \hat G}$:
\begin{equation}
  {\mathcal L}_{imp} = -\frac{1}{4} {\bf \hat G}^T_{\mu \nu} K {\bf \hat G}^{\mu
    \nu} + \frac{1}{2}{\bf \hat G}^T_{\mu} \hat M^2_{\bf \hat G}  {\bf \hat G}^\mu -
  \sum_{\psi^\prime} \bar \psi^\prime {\bf \hat l}^T 
  {\hat{\slashed{\bf G}}}
  \psi^\prime,
  \label{Limp}
\end{equation}
where objects with two Lorentz indices indicate the field strength, $K$
is a matrix encoding the kinetic mixing and $M^2_{\bf \hat{G}}$ is the 
neutral gauge boson mass squared matrix
\begin{equation}
  K:=\begin{pmatrix}
  1 & 0 & \sin \chi \\
  0 & 1 & 0 \\
  \sin \chi & 0 & 1 \\
  \end{pmatrix},
  \qquad
  M^2_{\bf \hat G}:= \begin{pmatrix}
  {g'}^2 v^2/4 & -gg' v^2/ 4 & 0 \\
  -gg' v^2/4 & \hat g^2v^2/4 & 0 \\
  0 & 0 & M_X^2 \\
  \end{pmatrix}.
\end{equation}
$M^2_{\bf \hat G}$ comes about as a consequence of the SM Higgs boson doublet field H acquiring a vacuum
expectation value $\langle H \rangle = (0, v/\sqrt{2})^T$.
We have parameterised the kinetic mixing terms by $\sin \chi$ in order to
enforce that its magnitude is less than unity: otherwise, one of the
eigenstates would have an incorrect kinetic term sign in the
Lagrangian.
We have also introduced the 3-vector
\begin{equation}
  {\bf \hat l}:= \left(g' \hat Y,\  g \hat L,\ g_X \hat X\right)^T \label{l}
  \end{equation}
where $g'$ is the hypercharge gauge coupling, $g$ the $SU(2)$ gauge coupling
and $g_X$ the $U(1)_X$ gauge coupling. 
The sum in (\ref{Limp}) is over the left and right-handed Weyl fermionic gauge
eigenstate representations of the model, $\psi^\prime$. 
\begin{table}
  \begin{center}
  \begin{tabular} {|c|cccccc|} \hline
    Field $\psi$ & $q_i^\prime=(u_{L_i}^\prime,\ d_{L_i}^\prime)^T$ &
    $l_i^\prime=(e_{L_I}^\prime,\ \nu_{L_i}^\prime)$ & $e_i^\prime$ & $d_i^\prime$ &
    $u_i^\prime$ & $\nu_i^\prime$ \\ \hline
    $SU(3)$      & 3 & 1 & 1 & 3 & 3 & 1 \\
    $SU(2)$      & 2 & 2 & 1 & 1 & 1 & 1 \\
    $Y_\psi$     & 1/6   & -1/2  & -1    & -1/3  & 2/3   & 0 \\    
    $X_\psi$ & $\delta_{i3}$ & -3$\delta_{i2}$ & -3$\delta_{i2}$ &  $\delta_{i3}$ & $\delta_{i3}$ & -3$\delta_{i2}$ \\
    \hline
  \end{tabular}
  \end{center}  
  \caption{Representation of gauge eigenstates of Weyl fermion fields under
    the $SU(3)\times SU(2)
  \times U(1)_Y \times U(1)_X$ gauge symmetry of the $B_3-L_2$ model.  \label{tab:charges}}
\end{table}
In (\ref{l}), the hat denotes an operator, thus
$\hat Y \psi^\prime = Y_{\psi^\prime} \psi^\prime$,
$\hat X \psi^\prime = X_{\psi^\prime} \psi^\prime$ and $\hat L$ annihilates
(anti-)right-handed Weyl fermionic fields or returns an eigenvalue of $T_3/2$
on an $SU(2)$ doublet field,  where $T_3$ is the diagonal
generator of the Lie algebra of $SU(2)$ (one half of the third Pauli matrix in
our conventions, in $SU(2)$
fundamental index space). 

The kinetic terms are diagonalised by the similarity transform
$P_\chi^{-1} K P_\chi$
with a non-unitary invertible matrix
\begin{equation}
  P_\chi := \begin{pmatrix}
    1 & 0 & - \tan_\chi \\
    0 & 1 & 0 \\
    0 & 0 & 1/\cos_\chi\\
    \end{pmatrix}.
  \end{equation}
$\hat M^2_{\bf \hat G}$ is diagonalised by $O_w^T \hat M^2_{\bf \hat G} O_w$ with
the orthogonal matrix
\begin{equation}
O_{\hat w} := \begin{pmatrix}
   \hat c_w  & -\hat s_w & 0 \\
   \hat s_w  &  \hat c_w & 0 \\
    0 & 0 & 1\\
    \end{pmatrix},
  \end{equation}
yielding eigenvalues 0 (with the photon field $A_\mu$ as its eigenvector),
$M_{\hat Z}^2:=v^2 (g^2+{g'}^2)/4$ (we call the corresponding eigenvector $\hat
Z_\mu$), and 
$M_X^2$, whose eigenvector is $X_\mu$. Here, $\hat c_w:=\cos
\hat \theta_w$ and $\hat s_w:=\sin \hat \theta_w$. $\hat \theta_w$ is the
would-be weak mixing angle in the kinetically-mixed basis, distinct from the
measured weak mixing angle $\theta_w$. 
The relationship between $\theta_w$ and $\hat \theta_w$  will be detailed below.

For brevity, we shall use the notation $t_\chi:=\tan \chi$, $c_\chi:= \cos
\chi$ and $s_\chi:=\sin \chi$. Writing the neutral gauge fields in a new basis
${\bf H}_\mu:=O_{\hat w}^T P_\chi^{-1} {\bf \hat G}_\mu$, (\ref{Limp}) becomes
\begin{equation}
  {\mathcal L}_{imp} = -\frac{1}{4} {\bf H}^T_{\mu \nu} {\bf H}^{\mu
    \nu} + \frac{1}{2}{\bf H}^T_{\mu}  M^2_{\bf H} {\bf H} -
  \sum_{\psi^\prime} \bar \psi^\prime {\bf \hat l}^T  P_\chi O_{\hat w} \slashed{\bf
    H} \psi^\prime, \label{Limp2}   
\end{equation}
where
\begin{equation}
  M_{\bf H}^2 :=
  \begin{pmatrix}
    0 & 0 & 0 \\
    0 & M_{\hat Z}^2 & M_{\hat Z}^2 \hat s_w t_\chi \\
    0 & M_{\hat Z}^2 \hat s_w t_\chi & M_X^2 / c_\chi^2 + M_X^2 \hat s_w^2
    t_\chi^2 \\
    \end{pmatrix}
\end{equation}
which is diagonalised by a similarity transform 
$O_z^T M_{\bf H}^2 O_z$
with the orthogonal matrix
\begin{equation}
  O_z :=
  \begin{pmatrix}
    1 & 0 & 0 \\
    0 & c_z & -s_z \\
    0 & s_z & c_s \\
    \end{pmatrix}
\end{equation}
provided that
\begin{equation}
  \tan 2 \theta_z = \frac{-2 M_{\hat Z}^2 \hat s_w s_\chi c_\chi}{M_X^2 +
    M_{\hat Z}^2 (\hat s_w^2 s_\chi^2 - c_\chi^2)}. \label{thetaz}
  \end{equation}
We define the propagating (or `physical') eigenstates
${\bf P}_\mu:=(A_\mu,\ Z_\mu,\ Z^\prime_\mu)^T:= O_z^T O_{\hat w}^T P_\chi^{-1} {\bf \hat G}_\mu$ to write
\begin{equation}
  {\mathcal L}_{imp} = -\frac{1}{4} {\bf P}^T_{\mu \nu} {\bf P}^{\mu\nu}
  +\frac{1}{2} {\bf P}_\mu M^2_{\bf P} {\bf P}^\mu
  - \sum_{\psi^\prime} \bar \psi^\prime {\bf \hat l}^T C \slashed{\bf P} \psi^\prime,
  \label{penultL}
  \end{equation}
where
\begin{equation}
  C:=O_{\hat w}^T P_\chi O_{\hat w} O_z
\end{equation}
and $M_{\bf P}^2=\text{diag}(0, M_Z^2, M_{Z^\prime}^2)$, where $M_Z$ and
$M_{Z^\prime}$ are the physical $Z^0$ and $Z^\prime$ boson masses,
respectively.  

The physical parameters are related to the model´s parameters
by~\cite{Cheng:2024hvq} 
\begin{eqnarray}
  M_{\hat Z}^2 &=& M_Z^2 \left[1 + s_z^2
    \left(\frac{M_{Z^\prime}^2}{M_Z^2}-1\right) \right], \label{mzhat} \\
  M_{X}^2 &=& \frac{c_\chi^2}{1+\hat s_w^2 s_\chi^2} \left[ s_z^2 M_Z^2 +
    c_z^2 M_{Z^\prime}^2 \right], \\
  M_{\hat Z} \hat c_w \hat s_w &=& M_Z c_w s_w \label{swhat}
\end{eqnarray}
along with (\ref{thetaz}).
$c_w$ is the cosine of the weak mixing
angle measured by experiments; $c_w = M_W/M_Z$ where $M_W$ is
the $W-$boson mass. 
We wish to solve the system of equations (\ref{thetaz}), (\ref{mzhat})-(\ref{swhat}) for a given $M_Z$ (taken from
experimental measurement), $M_{Z^\prime}$
and $s_\chi$. 
We shall solve these equations using approximation which is good enough for
our purposes here in
\S{\ref{sec:coup}} and detail an iterative procedure in order to obtain
an arbitrarily close approximation numerically in Appendix~\ref{sec:app}.

The final term in (\ref{penultL}) yields the coupling of the $Z^\prime$ to the
gauge eigenstates of the fermions $\psi^\prime$. In order to make contact with
phenomenology, we need to know the couplings of the $Z^\prime$ to the mass
eigenstates of the fermionic fields
\begin{equation}
  \psi_i := (V_{\psi}^\dag)_{ij} \psi^\prime_j,
  \end{equation}
where $V_{\psi}$ is a 3 by 3 unitary matrix in family space and $i,j,\in
\{1,2,3\}$ are family indices (with an implicit Einstein summation
convention). 
Following Ref.~\cite{Allanach:2024ozu}, we assume that, to a good approximation,
$V_{l_L}=V_{e_R}=V_{u_R}=V_{d_R}=I_3$, the 3 by 3 identity matrix, whereas
\begin{equation}
  V_{d_L}:=
  \begin{pmatrix}
    1 & 0 & 0 \\
    0 & c_{sb} & s_{sb} \\
    0 & -s_{sb} & c_{sb} \\
    \end{pmatrix}.
  \end{equation}
$\theta_{sb}\neq 0$ will then facilitate tree-level $b \rightarrow s$
transitions via 
$Z^\prime$ couplings. We then obtain $V_{u_L}=V_{CKM}^\dag V_{d_L}$ and 
$V_{\nu_L}=V_{PMNS}^\dag$, where $V_{CKM}$ is the Cabbibo-Kobayashi-Maskawa
matrix and $V_{PMNS}$ is the Pontecorvo-Maki-Nakagawa-Sakata
matrix~\cite{ParticleDataGroup:2024cfk}. The assumptions above about
the 
$V_\psi$ are strong, but have the consequence that strong constraints from
some flavour changing $Z^\prime$ couplings are evaded (for example from
charged lepton flavour violating processes).

The final term in (\ref{penultL}) can be expanded in terms of the mass
eigenstates in order to calculate 
the $Z^\prime$ couplings in the malaphoric $B_3-L_2$ model,
on which the phenomenology of the model heavily depends.
We shall perform this expansion in \S\ref{sec:coup}. 
In particular, 
a family universal part coming from the kinetic mixing will provide $Z^\prime$
couplings to the first two 
families of quark, enhancing the LHC production cross section.
We shall then analytically solve  
(\ref{thetaz}),(\ref{mzhat})-(\ref{swhat})
approximately.
In \S\ref{sec:cross}, we detail how we use
the ATLAS di-lepton searches to place bounds upon the
parameter space of the model. The results of this are then shown
before concluding in \S\ref{sec:conc}.

\section{$Z^\prime$-fermion couplings in the malaphoric $B_3-L_2$
  model\label{sec:coup}}
From now on, we shall write a 3-vectors of fermion fields in family space in
bold: ${\bm \psi}:=(\psi_1,\ \psi_2,\ \psi_3)^T$.
Expanding the right-most term of (\ref{penultL}) in terms of the mass eigenstates 
of these, we obtain the coupling of the $Z^\prime$ bosonic
field to them
\begin{eqnarray}
    {\mathcal L}_{\bar \psi Z^\prime \psi} &=& 
  -\overline{\bm u_L} \left(
  g_X \frac{c_z}{c_\chi} \Lambda_\xi^{u_L} + g^{u_L} I_3
  \right) \slashed{Z}^\prime {\bm u_L}  
  -\overline{\bm d_L} \left(
  g_X \frac{c_z}{c_\chi} \Lambda_\xi^{d_L} + g^{d_L} I_3
  \right) \slashed{Z}^\prime {\bm d_L} \nonumber \\ &&
 -\overline{\bm u_R} \left(
  g_X \frac{c_z}{c_\chi} \Lambda_\xi^{u_R} + g^{u_R} I_3
  \right) \slashed{Z}^\prime {\bm u_R} 
 -\overline{\bm d_R} \left(
  g_X \frac{c_z}{c_\chi} \Lambda_\xi^{d_R} + g^{d_R} I_3
  \right) \slashed{Z}^\prime {\bm d_R} \nonumber \\ &&
  -\overline{\bm \nu_L} \left(-3
  g_X \frac{c_z}{c_\chi} \Lambda_\Xi^{\nu_L} + g^{\nu_L} I_3
  \right)
  \slashed{Z}^\prime {\bm \nu_L}  
  -\overline{\bm e_L} \left(-3
  g_X \frac{c_z}{c_\chi} \Lambda_\Xi^{e_L} + g^{e_L} I_3
  \right) \slashed{Z}^\prime {\bm e_L} \nonumber \\ &&
  +3 \overline{\bm \nu_R} 
  g_X \frac{c_z}{c_\chi} \Lambda_\Xi^{\nu_R} 
  \slashed{Z}^\prime {\bm \nu_R}    
  -\overline{\bm e_R} \left(-3
  g_X \frac{c_z}{c_\chi} \Lambda_\Xi^{e_R} + g^{e_R} I_3
  \right) \slashed{Z}^\prime {\bm e_R},
  \end{eqnarray}
where $\Lambda_\xi^{\psi}:=V_\psi^\dag \xi V_\psi$,
\begin{equation}
  \xi := \begin{pmatrix}
    0 & 0 & 0 \\
    0 & 0 & 0 \\
    0 & 0 & 1 \\
  \end{pmatrix}, \qquad
  \Xi := \begin{pmatrix}
    0 & 0 & 0 \\
    0 & 1 & 0 \\
    0 & 0 & 0 \\
  \end{pmatrix},
\end{equation}
and the $g^\psi$ are the family universal components of the $Z^\prime$
coupling to fermions:
\begin{eqnarray}
  g^{u_L}&:=&\hat g_Z \left( \frac{1}{2} - \frac{2}{3} \hat s_w^2 \right)
  \left(c_z \hat s_w t_\chi - s_z \right) - \frac{2}{3} e c_z \hat c_w
  t_\chi, \nonumber \\
  g^{d_L}&:=&\hat g_Z \left( -\frac{1}{2} + \frac{1}{3} \hat s_w^2 \right)
  \left(c_z \hat s_w t_\chi - s_z \right) + \frac{e}{3} c_z \hat c_w
  t_\chi, \nonumber \\
  g^{u_R}&:=&\hat g_Z \left( - \frac{2}{3} \hat s_w^2 \right)
  \left(c_z \hat s_w t_\chi - s_z \right) - \frac{2}{3} e c_z \hat c_w
  t_\chi, \nonumber \\
  g^{d_R}&:=&\hat g_Z \frac{\hat s_w^2}{3}
  \left(c_z \hat s_w t_\chi - s_z \right) + \frac{e}{3}  c_z \hat c_w
  t_\chi, \nonumber \\   
  g^{\nu_L}&:=& \frac{\hat g_Z}{2}
  \left(c_z \hat s_w t_\chi - s_z \right), \nonumber \\
  g^{e_L}&:=&\hat g_Z \left( -\frac{1}{2} + \hat s_w^2 \right)
  \left(c_z \hat s_w t_\chi - s_z \right) - e c_z \hat c_w
  t_\chi, \nonumber \\
  g^{e_R}&:=&\hat g_Z \hat s_w^2 
  \left(c_z \hat s_w t_\chi - s_z \right) + e c_z \hat c_w
  t_\chi,
\end{eqnarray}
where $\hat g_Z := e / (\hat s_w \hat c_w)$ and 
$e$ is the electromagnetic gauge coupling. 

As we mention in \S\ref{sec:intro}, the malaphoric $B_3-L_2$ was fit
 via the SMEFT
to
various measurements in Ref.~\cite{Allanach:2024ozu}: $b \rightarrow s$ transition
observables and di-lepton scattering cross sections\footnote{Throughout the rest of the article, this is the `fit'
referred to.}. 
The SMEFT was expanded in powers of $g_X / M_X$ and the 95$\%$ CL region of
good fit was phrased in terms of $\{g_X/M_X,\ s_\chi/M_X,\ \theta_{sb}\}$.
The di-lepton bump hunt at the LHC is phrased in terms of $M_{Z^\prime}$
however and we must solve (\ref{thetaz}) and (\ref{mzhat})-(\ref{swhat}) for
it. 
$M_Z$ has been measured with high precision and so we take it as an
input here (we shall fix it to its experimental central value), as we do
$s_w$. The final list of input parameters is then $\{ M_Z,\ s_w,\ s_\chi,\
M_X\}$ whereas we need to solve the equations for
$\{ \theta_z,\ \hat s_w,\ M_{\hat Z},\ M_{Z^\prime}\}$.
We know of no exact analytic solution in
closed form. 

We shall make use of the fit region derived in Ref.~\cite{Allanach:2024ozu}
from the SMEFT given by integrating out the $Z^\prime$.
Since the fit region used, among other observables at lower scales, cross-section measurements
from LEP2 at a centre of mass energy of up to 206.5 GeV encoded via the SMEFT,
SMEFT applicability
requires that $M_{Z^\prime}$ should be much larger than this energy
scale. 
This implies that we may use
$\epsilon:=(M_Z/M_{Z^\prime})^2\ll 1$ as a perturbative expansion parameter in
order to find an approximate solution. Here,
an approximate solution, valid up to a factor
$(1+\mathcal{O}(\epsilon))$, is
easily precise enough for the purpose of predicting LHC 
$Z^\prime$ production cross sections, which have uncertainties larger
than the order one percent (or less) uncertainty coming from the perturbation expansion: 
\begin{eqnarray}
  \theta_z &=& - \frac{M_Z^2}{M_{Z^\prime}^2} s_w t_\chi (1 + s_w^2 s_\chi^2),
  \label{approx1} \\
  \hat s_w &=& s_w,  \\
  M_{\hat Z} &=& M_Z, \\
  M_{Z^\prime} &=& \frac{\sqrt{1+ s_w^2 s_\chi^2}}{c_\chi} M_{X}. \label{approx2}
  \end{eqnarray}
We remark that (\ref{approx1})-(\ref{approx2}) provide a valid approximation for
any $-1 < s_\chi < 1$ and we note that $Z-Z^\prime$ mixing is small: $\theta_z
\sim {\mathcal{O}}(\epsilon)$.
The malaphoric $B_3-L_2$ model described here
has been encoded (neglecting the flavon field $\theta$)
in {\tt UFO}~\cite{Degrande:2011ua} format using
(\ref{approx1})-(\ref{approx2}) 
and made 
available in the ancillary directory of the {\tt arXiv} preprint of this
paper. 

(\ref{thetaz}) and (\ref{mzhat})-(\ref{swhat}) could instead be solved
numerically to arbitrary precision by a  fixed point  
iteration method, as we detail in Appendix~\ref{sec:app}. A more precise
method such 
as the one in Appendix~\ref{sec:app} could be useful in the future if
(for example) one wished to use very precise experimental data from a 
future high energy $e^+ e^-$ collider such as the FCC-ee~\cite{FCC:2018evy} or 
CEPC~\cite{CEPCStudyGroup:2023quu} to bound the model.

\section{LHC $Z^\prime$ production cross-section and fit
  region \label{sec:cross}}
The re-casting of a 139 fb$^{-1}$ 13 TeV ATLAS search~\cite{ATLAS:2019erb} for $pp$ to resonant
di-lepton production 
follows that of Ref.~\cite{Allanach:2024jls} but we shall describe it in brief
here. ATLAS observed no significant excess on SM backgrounds and provides an observed
95$\%$ upper bound on $s(z, M_{Z^\prime})$, the total production cross-section
multiplied by 
branching ratio of a new physics state decaying to a specified flavour of
di-lepton 
pair (either di-electrons or di-muons). Here, $z$ is defined to be the new
physics state's resonance width  
divided by its mass. ATLAS provides the bounds upon both a narrow width $s(0,
M_{Z^\prime})$ and a resonance whose width is a tenth of its mass, $s(0.1,
M_{Z^\prime})$. The function
\begin{equation}
  s(z, M_{Z^\prime}) = s(0, M_{Z^\prime}) \left(\frac{s(0.1,
    M_{Z^\prime})}{s(0,M_{Z^\prime})} \right)^{z/10}  \label{interp}
\end{equation}
has been demonstrated to provide a good approximation for some other values
of $z$ between 0 and 0.1 that were also used by ATLAS for the di-electron 
channel~\cite{Allanach:2024jls} as well as for the di-muon
channel~\cite{Allanach:2019mfl}.
We shall use a superscript on $s$ to denote the lepton flavour,
i.e.\ $s^{\mu^+\mu^-}$ or $s^{e^+e^-}$.
Here, we shall use (\ref{interp}) to
interpolate for $0\leq z \leq 1$
and to extrapolate for $z > 0.1$, although if there is a region of
parameter space on a plot that has extrapolation, we shall demarcate it
the plot in question, since validation of (\ref{interp})
has not been performed there. 

As mentioned above, the malaphoric $B_3-L_2$ model was fit via the SMEFT to LEP2 data,
to $b \rightarrow s$ transition data and electroweak precision observables in
Ref.~\cite{Allanach:2024jls}. The fit depends, to a good approximation, on
only three effective
beyond-the-SM parameters which can be taken to be $(\sin \chi/M_X)$,
$(g_X/M_X)$ and 
$\theta_{sb}$. On the other hand, the collider physics depends sensitively
on $\sin \chi$, $M_X$ and $g_X$ but in practice, not sensitively on $\theta_{sb}$. In
the parameter region considered, $\sin \theta_{sb}$ is not
close to unity~\cite{Allanach:2024ozu}, where the $s \bar s \rightarrow
Z^\prime$ channel could change 
appreciably and alter the production cross-section.
\begin{table*}
  \begin{center}
    \begin{tabular}{|c|c|c|} \hline
                  &  $\theta_{sb}=-0.19$ & $\theta_{sb}=0$ \\ \hline
  $M_{Z^\prime}$/TeV &  4.184              & 4.184 \\
      $\Gamma_{Z^\prime}$/GeV & 30.5           & 30.5 \\
      $BR(Z^\prime \rightarrow e^+ e^-)$ & 0.01 & 0.01 \\
      $BR(Z^\prime \rightarrow \mu^+ \mu^-)$ & 0.48 & 0.48 \\
      $BR(Z^\prime \rightarrow t \bar t)$ & 0.20 & 0.21 \\
      $BR(Z^\prime \rightarrow b \bar b)$ & 0.08  & 0.08 \\
      $\sigma(u \bar u \rightarrow Z^\prime \rightarrow \mu^+ \mu^)$/fb &
   0.034   & 0.034 \\
      $\sigma(d \bar d \rightarrow Z^\prime \rightarrow \mu^+ \mu^)$/fb &
   0.0032   & 0.0032 \\
      $\sigma(c \bar c \rightarrow Z^\prime \rightarrow \mu^+ \mu^)$/fb &
   3.2$\times 10^{-4}$ & 3.1$\times 10^{-4}$ \\
   $\sigma(s \bar s \rightarrow Z^\prime \rightarrow \mu^+ \mu^)$/fb &
   1.4$\times 10^{-4}$ & 1.1$\times 10^{-4}$ \\
   $\sigma(b \bar b \rightarrow Z^\prime \rightarrow \mu^+ \mu^)$/fb &
   6.2$\times 10^{-4}$ & 6.5$\times 10^{-4}$ \\
   $\sigma(p p \rightarrow Z^\prime b \rightarrow \mu^+ \mu^- b)$/fb &
   3.6$\times 10^{-5}$& 3.4$\times 10^{-5}$\\
   $\sigma(p p \rightarrow Z^\prime \rightarrow e^+ e^-)$/fb & 9.1$\times
   10^{-4}$ & 9.1$\times 10^{-4}$ \\
   $s^{\mu^+\mu^-}$ & 0.048 & 0.048 \\
   $s^{e^+ e^-}$ & 0.037 & 0.037 \\
 \hline      \end{tabular}
    \end{center}
  \caption{Example 13 TeV LHC $Z^\prime$ production cross-section
    contributions and other quantities of 
    interest for $M_X=4$~TeV, $s_\chi=-0.2$, $g_X=0.1$.
    We have neglected to list any cross-section contributions below the
    $10^{-5}$~fb level.\label{tab:point}}
\end{table*}
We estimate the LHC
production cross-section and $Z^\prime$ total width and decay rates using {\tt
  MadGraph2.9.21} at tree-level order~\cite{Alwall:2014hca}. 
We pick an example point in parameter space from Ref.~\cite{Allanach:2024jls} which is just within the 95$\%$
fit window and show various quantities of
interest  in Table~\ref{tab:point}.
The non-zero $\theta_{sb}$ chosen is taken from its best-fit value, given the other parameters chosen. 
From the table, we see by comparing the numerical values of the two
rightmost columns that changing $\theta_{sb}$ does not change the branching
ratios or $Z^\prime$ production cross-section times branching ratio much.
This qualitative statement holds over the rest of the parameter space that
we shall cover.
From now on in the present paper, in our 
estimates for the cross-section, we shall approximate $\theta_{sb}=0$ when
calculating the limits coming from LHC resonant di-lepton production.

We see from the table that the partonic channel dominating the $Z^\prime$
production cross-section is via $u \bar u\rightarrow Z^\prime$: this proceeds via the hypercharge
mixing and is an order of magnitude larger than the $d \bar d$ contribution
because of a combination of parton distribution functions and the larger
hypercharge of $u_R$ quarks (implying a larger $Z^\prime$ coupling to them via
the kinetic mixing mixing). We also see that the associated production $pp \rightarrow Z^\prime b$ is
small. This associated production cross-section can be of a similar order to
that of
direct $p p 
\rightarrow Z^\prime$ in the case where the $b \bar b$ production channel
dominates, as it did in the original, kinetically unmixed $B_3-L_2$ model. In
that case, a CMS analysis~\cite{CMS:2023nzs} searching for resonant di-muons plus
$b-$jets can be used to acquire extra sensitivity. Here though, it should only
be relevant near the $s_\chi \rightarrow 0$ limit, which (as we shall see
below) is outside the region of good fit to flavour and other data. We shall
therefore concentrate on the inclusive resonant di-lepton production
cross-sections.

\begin{figure*}
  \begin{center}
    \unitlength = 0.9\textwidth
    \begin{picture}(1,1)(0,0)
    \put(-0.2,0.5){\includegraphics[width=0.765 \textwidth]{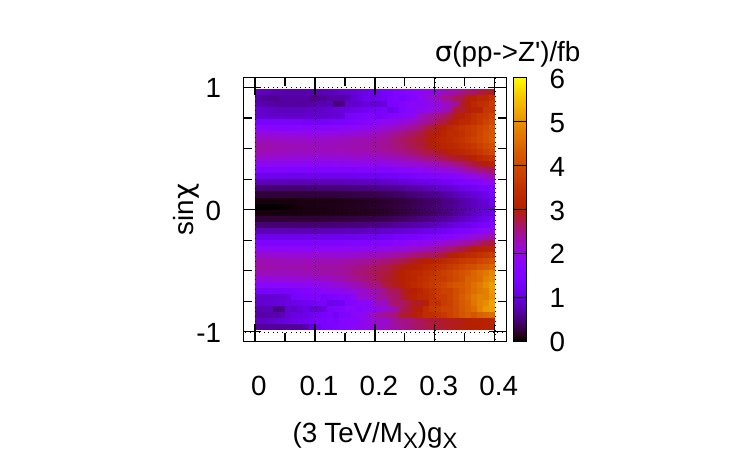}}
    \put(0.3,0.5){\includegraphics[width=0.765\textwidth]{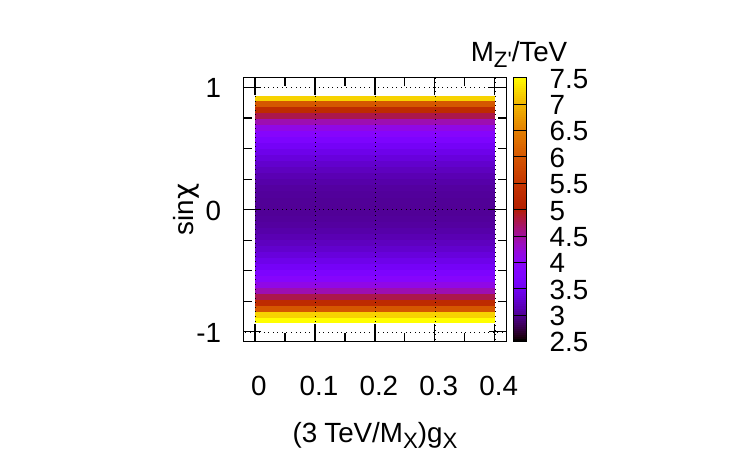}} 
    \put(-0.2,0){\includegraphics[width=0.765\textwidth]{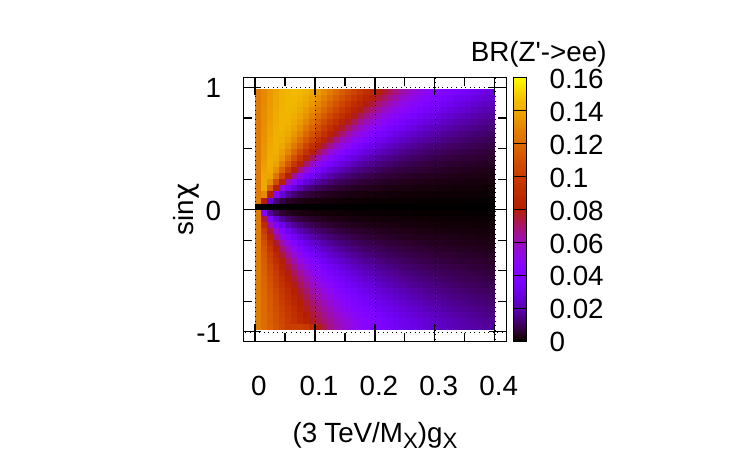}}
    \put(0.3,0){\includegraphics[width=0.765\textwidth]{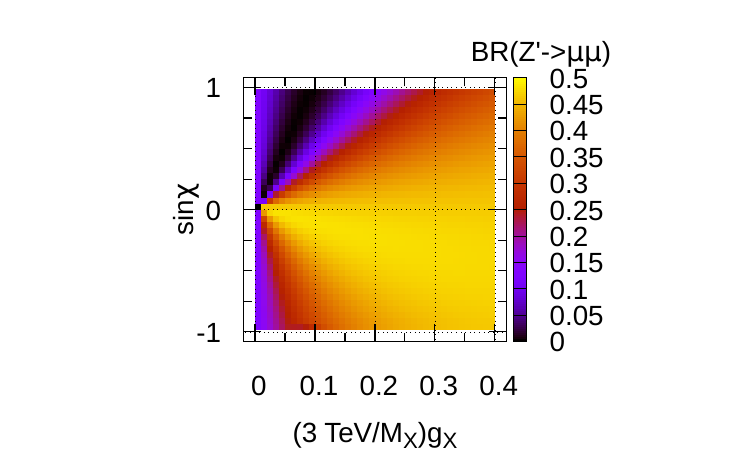}}
    \put(0,0.92){(a)}
    \put(0.5,0.92){(b)}
    \put(0,0.45){(c)}
    \put(0.5,0.45){(d)}        
    \end{picture}
  \end{center}
  \caption{Properties of the parameter space of the malaphoric $B_3-L_2$ model
    at $M_X=3$ 
    TeV. (a) displays the $Z^\prime$ production cross-section at the 13 TeV
    LHC. (b), (c) and (d) show various
  $Z^\prime$ properties as labelled on the colour bar. \label{fig:3TeVparam}}
\end{figure*}
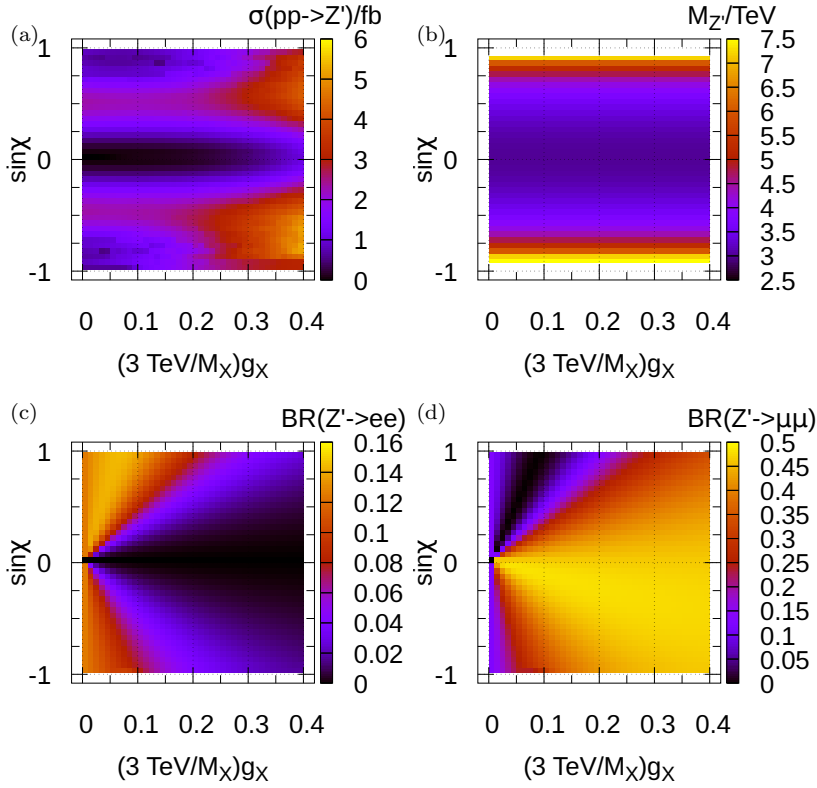
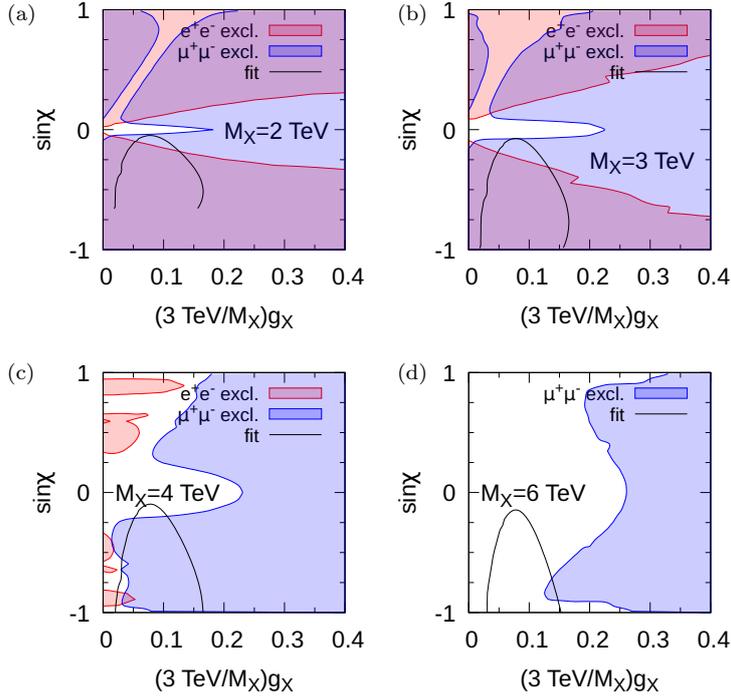
\begin{figure*}
  \begin{center}
    \unitlength = 0.9\textwidth
    \begin{picture}(1,0.9)(0,0.1)
    \put(-0.15,0.5){\includegraphics[width=0.7 \textwidth]{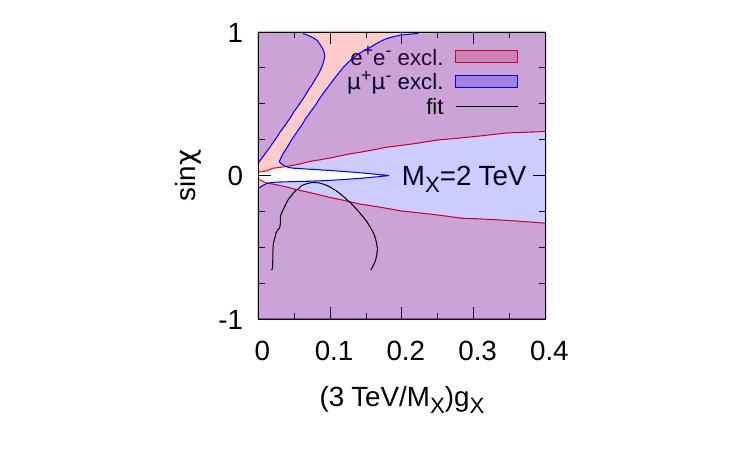}}
    \put(0.3,0.5){\includegraphics[width=0.7\textwidth]{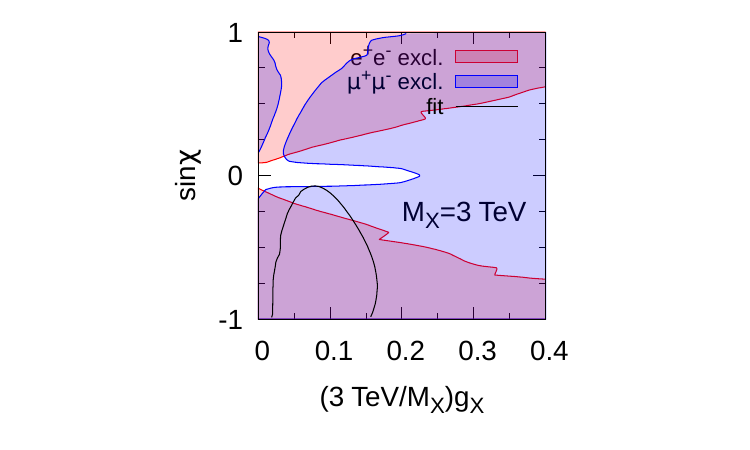}} 
    \put(-0.15,0.05){\includegraphics[width=0.7\textwidth]{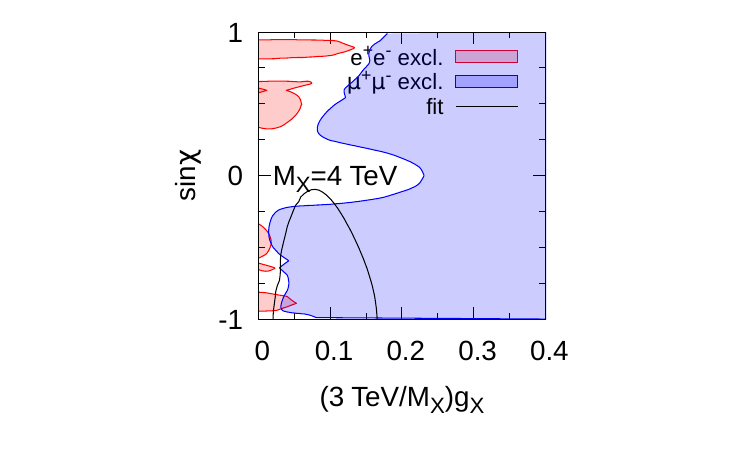}}
    \put(0.3,0.05){\includegraphics[width=0.7\textwidth]{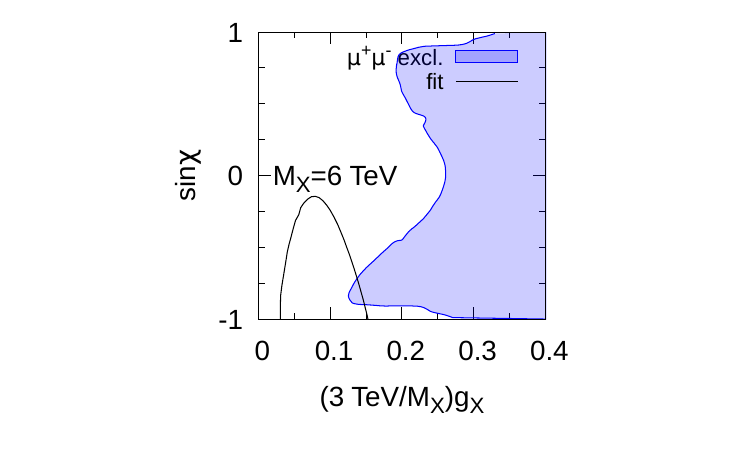}}
    \put(0,0.92){(a)}
    \put(0.48,0.92){(b)}
    \put(0,0.475){(c)}
    \put(0.48,0.475){(d)}        
    \end{picture}
  \end{center}
  \caption{Parameter space of the malaphoric $B_3-L_2$ model at various values
    of $M_X$. Regions that are excluded at 95$\%$ CL~\cite{ATLAS:2019erb} are shown by the colour
    in the legend. The region bounded within each black contour is preferred
    by fits to 
    data including $b \rightarrow s l^+l^-$ at the 95$\%$ CL, from Ref.~\cite{Allanach:2024ozu}. \label{fig:XTeVbounds}}
\end{figure*}

To begin with, we fix $M_X=3$ TeV and vary $\sin \chi$ and
$g_X/M_X$ (i.e.~effectively varying $g_X$).
We have picked the domain of the abscissa to
encompass the region of good fit, as we shall see in Fig.~\ref{fig:XTeVbounds}.
The 13 TeV LHC $Z^\prime$ total production cross section is shown in
Fig.~\ref{fig:3TeVparam}a. 
The $Z^\prime$ total production cross section increases towards the right-hand
side of the plot, where the gauge coupling is larger.
It becomes small when $s_\chi\rightarrow 0$ because there the $u \bar u$ coupling
to the $Z^\prime$ vanishes.
$M_{Z^\prime}$ 
varies across this two-parameter plane according to (\ref{approx2}) and is
shown in Fig.~\ref{fig:3TeVparam}b. We see that it is around 3 TeV except
towards $|\sin \chi|=1$, where $M_{Z^\prime}$ reaches up to 7 TeV. Near $s_\chi=1$, the
total cross-section is suppressed because of the effects of this large mass,
as a quick reference to Fig.~\ref{fig:3TeVparam}a confirms. 
The two relevant leptonic branching ratios
$BR(Z^\prime \rightarrow e^+ e^-)$ and $BR(Z^\prime \rightarrow \mu^+ \mu^-)$
are shown in Fig.~\ref{fig:3TeVparam}c and Fig.~\ref{fig:3TeVparam}d, respectively. Note that, to a good
  approximation, this model predicts that $BR(Z^\prime \rightarrow
  \tau\tau)=BR(Z^\prime \rightarrow ee)$.

\begin{figure*}
  \begin{center}
    \unitlength = 0.9 \textwidth
    \begin{picture}(1,1)(0,0)
    \put(-0.2,0.5){\includegraphics[width=0.765 \textwidth]{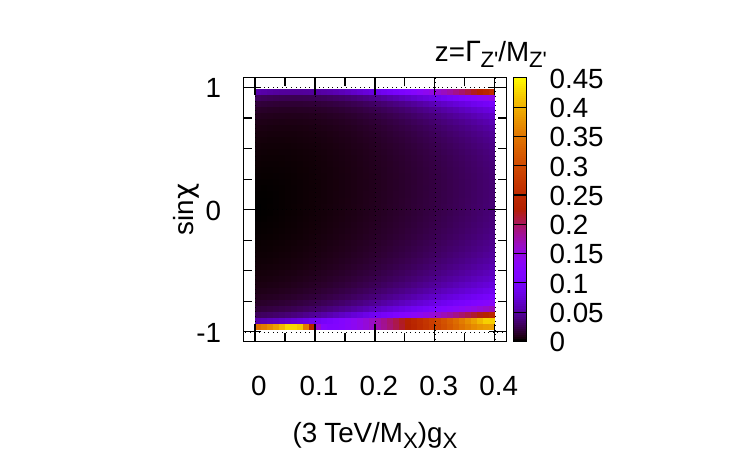}}
    \put(0.3,0.5){\includegraphics[width=0.765\textwidth]{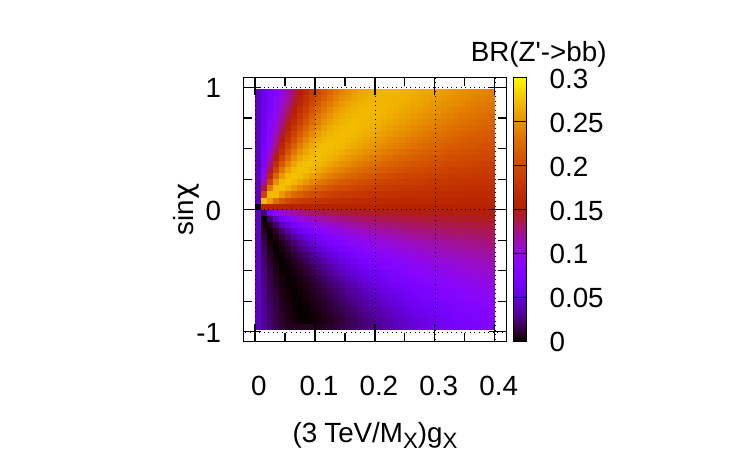}} 
    \put(-0.2,0){\includegraphics[width=0.765\textwidth]{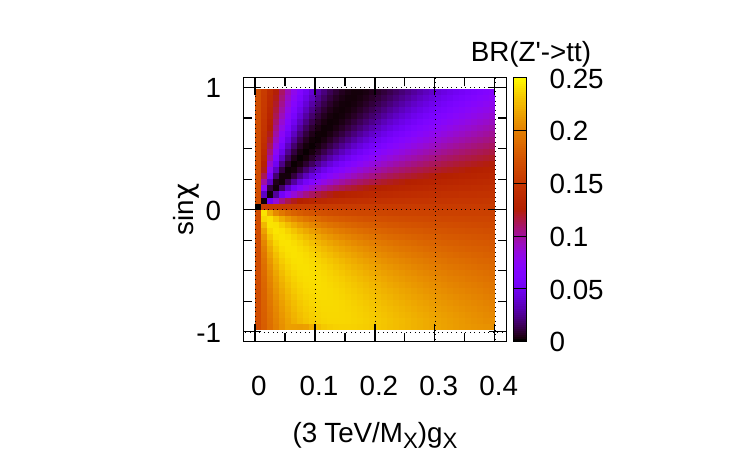}}
    \put(0.3,0){\includegraphics[width=0.765\textwidth]{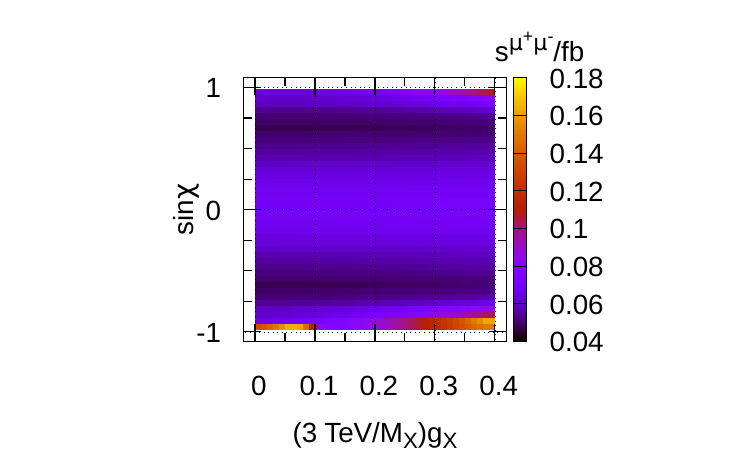}}
    \put(0,0.92){(a)}
    \put(0.5,0.92){(b)}
    \put(0,0.45){(c)}
    \put(0.5,0.45){(d)}    
    \end{picture}
  \end{center}
  \caption{$Z^\prime$ properties of the malaphoric $B_3-L_2$ model at $M_X=3$
    TeV. (a) shows the width of the $Z^\prime$ divided by its mass.
    (b) shows the branching ratio of the $Z^\prime$ into $b \bar b$ pairs,
    (c) shows the branching ratio into $t \bar t$ pairs and
    (d) shows the observed 95$\%$ CL upper bound on total $pp\rightarrow
    Z^\prime \rightarrow \mu^+\mu^-$
    cross-section derived from the 13 TeV ATLAS 139 fb$^{-1}$ resonant
    di-lepton search~\cite{ATLAS:2019erb}\label{fig:3TeVprop}}
\end{figure*}
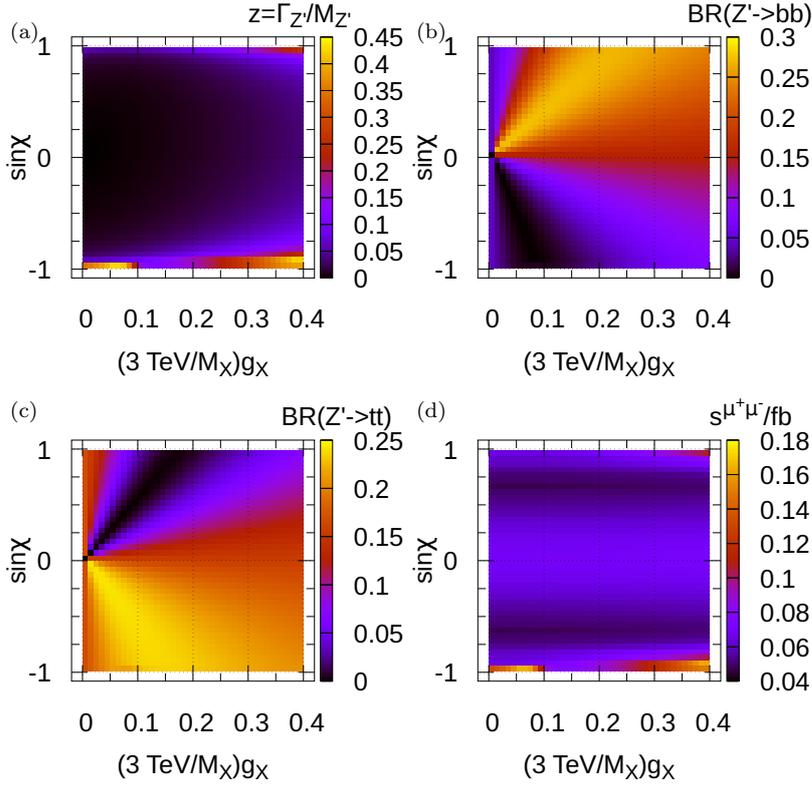
Two other branching ratios of the $Z^\prime$ (into $b \bar b$ and $t \bar t$
final states) and the total width of the $Z^\prime$ boson divided by its mass
are shown in Fig.~\ref{fig:3TeVprop}.  
The upper bounds on resonant $Z^\prime$ production followed by decay into
$\mu^+\mu^-$ coming from the resonant ATLAS di-lepton search over this
parameter space are functions
of the production cross-section, the relevant leptonic branching
ratio as well as $z$ and $M_{Z^\prime}$. 
The fraction of $Z^\prime$ width over its mass is displayed in the figure. The
fact that it is less than 0.45 means that nowhere in parameter space is the
resonance frequency becoming so wide that perturbation theory is obviously
breaking down (this happens near fractions of around 1).

We now turn to constraints upon the parameter space. We show these for
various different chosen values of $M_X$ in Fig.~\ref{fig:XTeVbounds}.
The white region on each plot is currently allowed at the 95$\%$ CL by the
ATLAS resonant di-lepton search~\cite{ATLAS:2019erb}. The region within the
black curve is preferred by the global fit to measurements of $b \rightarrow s$ observables,
electroweak parameters and LEP2 di-lepton production
cross-sections~\cite{Allanach:2024ozu}. 
The locus of this curve depends upon $M_X$: to a good approximation
(neglecting small renormalisation group effects between $M_X$ and
$M_{Z}$) it depends only on the ratio $\sin \chi/M_X$ and the abscissa.
In Ref.~\cite{Allanach:2024ozu}, the curve was determined for $M_X=3$ TeV for values of $\sin
\chi \leq 1$. For $M_X=2$ TeV, this translates into a curve at values of $\sin
\chi \leq 2/3$, explaining why the curve in Fig.~\ref{fig:XTeVbounds}a is
incomplete. From Figs.~\ref{fig:XTeVbounds}a and~\ref{fig:XTeVbounds}b, we
see that there the region of good fit is ruled out by the ATLAS resonant
di-lepton search (in particular in the di-muon channel) for $M_X=2$ TeV and
$M_x=3$ TeV, respectively. However, Figs.~\ref{fig:XTeVbounds}d and~\ref{fig:XTeVbounds}d show that $M_X=4$ TeV and $M_X=6$ TeV have some
allowed parameter space in the region of good fit. 

\begin{figure*}
  \begin{center}
    \includegraphics[width=0.95 \textwidth]{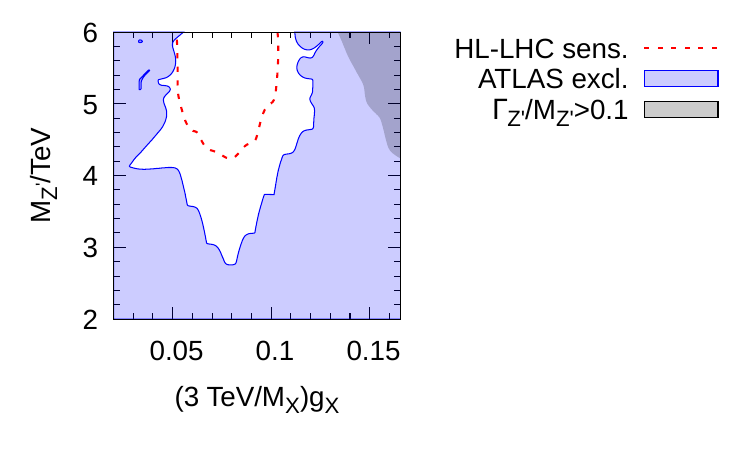}    
  \end{center}
  \caption{\label{fig:edge} ATLAS di-lepton resonance
    search exclusions in terms of the physical mass of the
    $Z^\prime$ boson $M_{Z^\prime}$~\cite{ATLAS:2019erb}; the
    coloured region 
    is excluded at the 95$\%$ CL whereas the dashed line shows the estimated
    HL-LHC sensitivity.
    In each case $\sin \chi$ has been adjusted to the value that gives the
    weakest constraint within the 95$\%$ fit region shown
    in Fig.~\ref{fig:XTeVbounds} (see the text for more detail).}
  \end{figure*}

Ideally, we wish to phrase the result of ATLAS searches in terms of the more
physical parameter $M_{Z^\prime}$. We do this by using a 3-dimensional scan
over $M_X$,  
$g_X$ and $\sin \chi$ within the 95$\%$ CL fit region. To effect this,
we scan over the black curve in Fig.~\ref{fig:XTeVbounds}b parameterised
jointly by the one-dimensional curve in 
$x:=(\text{3~TeV}/M_X)g_X$ and $y:=(\text{3~TeV}/M_X)\sin \chi$. $M_X$ is scanned between 2 TeV and 6 TeV. $\sin \chi$ is then scanned between the value
consistent with $y$ and\footnote{The parameter region $-1 <
\sin \chi < -0.95$ leads to values of $M_{Z^\prime}>6$ TeV, i.e.\ higher than
those relevant to the ATLAS bounds. As Fig.~\ref{fig:3TeVprop}a shows, this
region of 
parameter space also has wide $Z^\prime$
bosons, resulting from larger couplings, with less reliable perturbativity.} $-0.95$ in steps of
approximately\footnote{We scan $\sin \chi$ between $-0.95$ and the edge value 
$E=y M_X/\text{(3~TeV)}$ with $100(E-0.95)$ equally spaced steps, rounded down to the nearest integer.} 
$0.01$. 
For each of these values of $\sin \chi$, we compute
\begin{equation}
  R:=\text{max}\left\{ \sigma(pp \rightarrow Z^\prime \rightarrow \mu^+
  \mu^-)/s^{\mu^+\mu^-},\
  \sigma(pp \rightarrow Z^\prime \rightarrow e^+
  e^-)/s^{e^+e^-} \right\}. \label{R}
\end{equation}
In the plot, 
$\sin \chi$ is then fixed to the scanned value that returns the minimum value
of $R$. In this way, the approximate value of $\sin \chi$ is found which is
the least excluded by the ATLAS di-lepton searches. If we then plot the
contour $R=1$ across the $M_{Z^\prime}$ versus $(g_X/M_X)$ plane, we get a
picture
of the minimum amount of parameter space excluded (in other words, if we were
to move $\sin \chi$ away from this value, we could exclude more parameter
space).
We plot the resulting exclusion in Fig.~\ref{fig:edge}. We see that, within
the 95$\%$ fit r\'egime, one can infer that
$M_{Z^\prime}>2.8$ TeV from the Run II ATLAS resonant di-lepton search
whichever the value of the other parameters.
Next, we wish to provide an estimate of the HL-LHC sensitivity to the
model. For this, we re-purpose (\ref{R}), first changing $s^{\mu^+ \mu^-}$ and
$s^{e^+e^-}$ to the \emph{expected bound} rather than the observed one.
Since this may change the value of $\sin \chi$ which we use in the plot at
each point (remembering that this is fixed to the value that returns
the minimum value of $R$), the expected bound requires its own scan, which we perform.
We then scale the expected ATLAS sensitivity by the square
root\footnote{For some potential caveats associated with such a procedure, see
Ref.~\cite{Belvedere:2024wzg}.} of
the luminosity. We assume that the HL-LHC will acquire 3000 fb$^{-1}$
of integrated luminosity and so the expected excluded cross-section is
\begin{equation}
s^\text{HL-LHC}=s (z, M_{Z^\prime}) \sqrt{\frac{139~\text{fb}^{-1}}{3000 \text{~fb}^{-1}}}.
  \end{equation}
We see from Fig.~\ref{fig:edge} that our estimate yields that the HL-LHC
should be sensitive up to at least $M_{Z^\prime}=4.2$ TeV at 95$\%$ CL.

\section{Conclusions \label{sec:conc}}
The original $B_3-L_2$ model explained some $b \rightarrow s \mu^+ \mu^-$ measurements,
which were (prior to December 2022, unambiguously) in tension with SM
predictions. It predicts a TeV-scale neutral $Z^\prime$ vector boson, which decays
into di-muon pairs and could potentially be produced and found at the
LHC\@. Direct search bounds were rather weak ($M_{Z^\prime}>1.2$
TeV in the region of good fit~\cite{Allanach:2021gmj}) due to the fact that,
to a good approximation, the LHC production cross-section was dominated by $b \bar
b \rightarrow Z^\prime$ and so doubly suppressed by the $b-$quark parton
distribution function. 
Since then, the measurements of lepton flavour universality (LFU) ratios $R_K$
and $R_{K^\ast}$ have changed, requiring a change to the model such that new
physics effects in $b \rightarrow s e^+ e^-$ are also present. Here, we have
taken the suggestion that the original model has sizeable kinetic
mixing~\cite{Allanach:2024ozu}, and re-calculated LHC direct search bounds on 
the $Z^\prime$. In the new model, the $Z^\prime$ is dominantly produced at the
LHC by valence quark production, with a consequently much higher
cross-section. The bounds coming from direct searches at the LHC are then
expected to be stronger than that of the original model.
This is borne out by our analysis: ATLAS di-lepton searches at the 13 TeV LHC imply that the $Z^\prime$ boson of
the malaphoric $B_3-L_2$ model should have a mass of at least 2.8~TeV if
it is 
to explain the disparity between various current measurements of $b
\rightarrow s$ transitions in 
$B-$meson decays and their state-of-the-art SM predictions while remaining
compatible with measurements of electroweak 
observables and LEP2 di-lepton scattering measurements.
The CMS experiment has also performed
a similar di-lepton search~\cite{CMS:2021ctt}, with a similar
exclusion on cross-section multiplied by branching ratio to the ATLAS analysis
used in the present paper as a function of $M_{Z^\prime}$.
We expect bounds extracted from the CMS analysis to be very similar to those
presented here.

In terms of direct signatures, we have ignored processes involving the flavon
$\theta$. There is the possibility of a term in the potential $\lambda
\theta^\dag \theta H^\dag H$, which if $\lambda\neq0$ leads to Higgs-flavon
mixing. This would change Higgs decay and production rates~\cite{Allanach:2022blr}, as well as leaving open
the possibility of flavonstrahlung~\cite{Allanach:2020kss} via $p p
\rightarrow Z^\prime \theta$.
$\lambda$ is multiplicatively renormalised and therefore
the $\lambda=0$ limit is stable to
renormalisation corrections, so assuming that it is negligible is at least
self-consistent. $\lambda \neq 0$ in the unmixed $B_3-L_2$ model has been
studied in Ref.~\cite{Allanach:2022blr}.
We leave studies of the phenomenology of sizeable values of $\lambda$ in the 
kinetically-mixed case to future work.

We note that a simple tweak to the malaphoric $B_3-L_2$ model leads to another
possibility: one simply switches the 
$X$ charges of the third and second family leptons in the model. The
$Z^\prime$ coupling to the di-muon pairs would be entirely through the kinetic mixing
terms. The di-electron pair coupling to $Z^\prime$ would be identical to the
di-muon pair coupling, with the implication that 
lepton flavour should be 
universal between electrons and muons. The LFU observables
taken as a whole 
still prefer a disparity between the di-electron and di-muon channels at the
2$\sigma$ level~\cite{isidori} and so this option is currently somewhat
disfavoured.
Another model possibility is that $X:=3B_3-L_1-2L_2$, shown in Ref.~\cite{Allanach:2023uxz} to
fit flavour, electroweak and LEP2 measurements well. In that case, there are no
sizeable couplings to valence quarks and the current LHC bounds ($M_{Z^\prime}
> 1.2$ TeV) are consequently weaker
than those found in the present paper~\cite{Allanach:2024jls}.

One may wonder about the possibilities for ultra-violet model building in the
case with order unity kinetic mixing. We point out that the
malaphoric model is equivalent to a model \emph{with zero} kinetic mixing in
which the charge is instead 
assigned to be $X:=B_3-L_2 + \alpha Y$, where $\alpha \in \mathbb{Q}$ is
assigned appropriately (such charge assignments have
been proposed in the
literature~\cite{Altmannshofer:2019xda,Greljo:2021npi}). 

In the case that $M_{Z^\prime}>4.2$ TeV, one would require future colliders
to further probe the malaphoric $B_3-L_2$ model by resonant di-lepton searches, for example muon colliders
or FCC-hh~\cite{Allanach:2017bta,Azatov:2022itm}. The FCC-ee would provide diverse indirect
evidence via flavour tagging~\cite{Greljo:2024ytg}, which would help test the
model. 

\begin{acknowledgements}
We thank the Cambridge Pheno Working Group (especially N.\ Gubernari),
M.\ Goodsell and
P.\ Stangl for helpful discussions.
We thank CERN for hospitality extended while part of this work was
undertaken. The author would like to express special thanks to the Mainz
Institute for Theoretical Physics (MITP) of the CLuster of Excellence
PRISMA$^+$ (Project ID 390831469), for its hospitality and its partial support
during the completion of this work.
\end{acknowledgements}

\appendix

\section{Iterative solution of kinetic mixing parameter equations \label{sec:app}}
Here, we describe a method for the numerical solution of
(\ref{thetaz}), (\ref{mzhat})-(\ref{swhat}) to arbitrary 
precision by fixed point iteration. We begin with an approximation for $M_{Z^\prime}$,
$M_{\hat Z}$ and $\hat s_w$ and use the equations to derive a closer
approximation. We shall use a 
suffix $i$ on each of these quantities to denote the $i^{th}$ iteration.
The initial iteration ($i=0$) is fixed by the right-hand sides of our
approximate solution (\ref{approx1})-(\ref{approx2}).
First, we calculate
\begin{equation}
\theta_{z_i} = \frac{1}{2} \tan^{-1} \left(\frac{-2 c_\chi s_\chi \hat s_{w_i}
M^2_{\hat Z_i}}{M_{X}^2 + {M^2_{{\hat Z}_i}} (\hat s_{w_i}^2 s_\chi^2 - c_\chi^2)} \right). 
\end{equation}
This allows us to calculate the next iteration via 
\begin{equation}
M_{{\hat Z}_{i+1}} = M_Z \sqrt{1 + s_{z_i}^2 \left(\frac{M_{Z^\prime_i}^2}{M_Z^2} - 1\right)}, \label{mzhat_next}
\end{equation}
which is substituted into
\begin{eqnarray}
\hat s_{{2w_{i+1}}} &=& \frac{s_{2w} 
  M_Z}{M_{{\hat Z}_{i+1}}}, \nonumber \\
M_{Z^\prime_{i+1}} &=&
\frac{1}{c_{z_i} c_\chi}
\sqrt{M_X^2(1 + \hat s_{w_{i+1}}^2 s_\chi^2) - c_\chi^2 s_{z_i}^2 M_Z^2}
\end{eqnarray}
We find that this iterative procedure converges quickly: an example point
($M_{Z^\prime}=1$ TeV, $s_\chi=-0.86$) takes eight iterations to reach
a relative precision of order $10^{-16}$. The algorithm and example case is provided in the ancillary
directory associated with the {\tt arXiv} version of this paper in the form of
a {\tt python} program.

\bibliographystyle{JHEP-2}       
\bibliography{template}


\end{document}